\documentclass{amsart}
\usepackage{amsaddr}
\usepackage{amsmath}
\usepackage{amssymb}
\usepackage{gensymb}
\usepackage{breqn}
\usepackage{geometry}

\title{Comment on ``Inferring the Dynamics of Underdamped Stochastic Systems''}
\author{Yeeren I. Low}
\address{Department of Physics, University of Vermont, Burlington, Vermont 05405, USA}
\email{yeeren.low@uvm.edu}

\begin{document}
\begin{abstract}
	D.\ B.\ Br\"{u}ckner \textit{et al.}\ [\textit{Phys.\ Rev.\ Lett.}\ \textbf{125}, 058103 (2020)] have described a novel method for inferring the dynamics of systems governed by an underdamped Langevin equation in the presence of measurement noise. While this is a significant achievement, the paper also presents a number of significant errors. These are explained and corrected in this note.
\end{abstract}

\maketitle

The authors \cite{Bruckner} derive an estimator for the force projections in Eq.\ (S48). However, the magnitude of the ignored terms is incorrect. Inspecting Eq.\ (S43), extending the Taylor expansion to second order, one sees that the correction is not $\mathcal O(\Lambda)$ ($\Lambda$ being the measurement error covariance in Eq.\ (S38), except that the Dirac delta $\delta (t-t')$ should instead be Kronecker delta $\delta_{tt'}$) but rather $\mathcal O(\Lambda (\Delta t)^{-2})$. The evaluation of $\langle \hat c_{\alpha} \hat c_{\beta} \rangle$ also contains this correction. This is consistent with the requirement mentioned in the main text that the measurement error must be small compared to the typical displacement in a single time-step. The mistake is propagated in the subsequent equations. In the main text, it is also stated that the bias in Eq.\ (S45) is of order $\mathcal O((\Delta t)^{-3})$; however, as $\Lambda (\Delta t)^{-2}$ is required to be small, the bias is actually of order $\mathcal O((\Delta t)^{-1})$. Additionally, the second term on the r.h.s.\ of Eq.\ (S44) is of the same order as the ignored terms, thus the requirement that Eq.\ (S46) vanishes is of questionable relevance. As a result, the exact choice of $\overline {\mathbf y}$ seems to be of minor importance, and the claimed optimality of Eq.\ (S48b) appears not to be justified. (Incidentally, in the second term on the r.h.s.\ of Eq.\ (S48a), $\overline {\mathbf y}$ and $\hat {\mathbf w}$ should be $\tilde {\mathbf y}$ and $\check {\mathbf w}$, respectively, as defined in Eqs.\ (S72) and (S73).)

The second error concerns the estimation of the noise given in Eq.\ (S92). It is readily seen that Eq.\ (S70) (identical to Eq.\ (S92b)) cannot be correct, as the coefficients ought to sum to zero. This appears to be a typo as $-6$ should be $-3$ (derivation to follow). This error is not present in the Python code, so the numerical results are unaffected. However, this error is propagated to the subsequent steps in the derivation. I will now derive the corrected result. Let $\Delta \mathbf y^{(-/0/+)}$ be defined as in Eq.\ (S62), and define:
\begin{equation}
	\Delta^2 \mathbf y^{(-)} := \Delta \mathbf y^{(0)} - \Delta \mathbf y^{(-)}, \quad \Delta^2 \mathbf y^{(+)} := \Delta \mathbf y^{(+)} - \Delta \mathbf y^{(0)}.
\end{equation}
We shall seek estimators for $\sigma^2_{\mu \nu} (\Delta t)^3$ and $\Lambda_{\mu \nu}$ of the form:
\begin{equation} \label{eq:second-differences}
	k_0 \left({\Delta^2 y_{\mu}^{(-)} \Delta^2 y_{\nu}^{(-)} + \Delta^2 y_{\mu}^{(+)} \Delta^2 y_{\nu}^{(+)}}\right) + k_1 \left({\Delta^2 y_{\mu}^{(-)} \Delta^2 y_{\nu}^{(+)} + \Delta^2 y_{\mu}^{(+)} \Delta^2 y_{\nu}^{(-)}}\right),
\end{equation}
where $k_0$, $k_1$ are coefficients to be determined. We may write this as a linear combination of the $\Delta_{\mu \nu}^{(n,m)}$ defined in Eq.\ (S51). With the above choice, all quantities on the r.h.s.\ of Eq.\ (S53) which are linear in $n$ or $m$ will be eliminated when the linear combination is taken, due to vanishing of the second differences. (Incidentally, it appears that in Eq.\ (S53), $n$ and $m$ are swapped in some terms, and a factor of 1/2 is missing in $F_{\mu}$ from Eq.\ (S13), but these do not affect the final result.) Thus, of the terms written on the r.h.s.\ of Eq.\ (S53), only the second and third need to be considered. (Higher-order terms will be considered later.) We therefore have two coefficients to be determined for two estimators. For the estimator $\widehat {\sigma^2}_{\mu\nu} (\Delta t)^3$, this gives $k_0 = 6/11$ and $k_1 = 9/11$, while for the estimator $\hat \Lambda_{\mu \nu}$, this gives $k_0 = 1/44$ and $k_1 = -1/11$. These results have been previously given in \cite{low2024second} and are identical to Eqs.\ (S70) and (S71), except with the $-6$ replaced by $-3$ in Eq.\ (S70).

The third error concerns Eqs.\ (S92c) and (S92d) in the case of multiplicative noise, which were derived from the erroneous Eq.\ (S70). It has already been argued that the bias in the estimation of $\widehat {\sigma^2}_{\mu \nu \alpha}$ due to Eq.\ (S75) vanishes. Before discussing the bias for $\widehat {\sigma^2}_{\mu \nu \alpha}$, it should be mentioned that the neglected term in Eq.\ (S74) is of order $\mathcal O(\Lambda (\Delta t)^{-2})$, as has been explained in the context of the force estimator. This should be propagated to Eqs.\ (S77) and (S92a). There are two leading sources of bias for $\widehat {\sigma^2}_{\mu \nu \alpha}$ involving the measurement error when Eq.\ (S51) is used with Eq.\ (S50). One comes from the product of the force term with the measurement error, which is given by:
\begin{equation}
\frac {n^2 F_{\mu} \Delta \eta_{\nu}^{(m)} + m^2 F_{\nu} \Delta \eta_{\mu}^{(n)} } 2 (\Delta t)^2.
\end{equation}
Following Eq.\ (S77), the resulting bias is of order $\mathcal O(\Delta t, \Lambda (\Delta t)^{-2})$ and can thus be neglected, regardless of the choices of $a_n$ and $b_n$ in Eqs.\ (S72) and (S73). The other source of bias comes from the product of $\sigma_{\mu \rho} I_{0 \rho}^{(n)}$ with $\Delta \eta_{\nu}^{(m)}$. To evaluate the resulting bias, we substitute into Eq.\ (S74):
\begin{equation}
	\hat c_{\alpha} (\tilde {\mathbf x}, \check {\mathbf v}) = \hat c_{\alpha} (\mathbf x, \mathbf v) + [\partial_{x_{\mu}} \hat c_{\alpha} (\mathbf x, \mathbf v)] (\tilde x_{\mu} - x_{\mu}) + [\partial_{v_{\mu}} \hat c_{\alpha} (\mathbf x, \mathbf v)] (\check v_{\mu} - v_{\mu}) + \mathcal O(\Delta t),
\end{equation}
and similarly for $\partial_{x_{\mu}} \hat c_{\alpha} (\tilde {\mathbf x}, \check {\mathbf v})$ and $\partial_{v_{\mu}} \hat c_{\alpha} (\tilde {\mathbf x}, \check {\mathbf v})$. Expanding Eq.\ (S74) then gives:
\begin{equation}
	\begin{aligned}
		\hat c_{\alpha} (\tilde {\mathbf y}, \check {\mathbf w}) &= \hat c_{\alpha} (\mathbf x, \mathbf v) + [\partial_{x_{\mu}} \hat c_{\alpha} (\mathbf x, \mathbf v)] (\tilde x_{\mu} - x_{\mu}) + [\partial_{v_{\mu}} \hat c_{\alpha} (\mathbf x, \mathbf v)] (\check v_{\mu} - v_{\mu}) \\
		&\quad{} + \left\{{ \partial_{x_{\mu}} \hat c_{\alpha} (\mathbf x, \mathbf v) + [ \partial_{x_{\mu}} \partial_{x_{\nu}} \hat c_{\alpha} (\mathbf x, \mathbf v) ] (\tilde x_{\nu} - x_{\nu}) + [ \partial_{x_{\mu}} \partial_{v_{\nu}} \hat c_{\alpha} (\mathbf x, \mathbf v) ] (\check v_{\nu} - v_{\nu}) }\right\} (\tilde y_{\mu} - \tilde x_{\mu}) \\
		&\quad{} + \left\{{ \partial_{v_{\mu}} \hat c_{\alpha} (\mathbf x, \mathbf v) + [ \partial_{v_{\mu}} \partial_{x_{\nu}} \hat c_{\alpha} (\mathbf x, \mathbf v) ] (\tilde x_{\nu} - x_{\nu}) + [ \partial_{v_{\mu}} \partial_{v_{\nu}} \hat c_{\alpha} (\mathbf x, \mathbf v) ] (\check v_{\nu} - v_{\nu}) }\right\} (\check w_{\mu} - \check v_{\mu}) \\
		&\quad{} + \mathcal O(\Delta t, \Lambda (\Delta t)^{-2}).
	\end{aligned}
\end{equation}
From Eqs.\ (S21), (S23) (in which the superscripts (1) and (2) are swapped), (S72), and (S73), we see that the resulting bias is also of order $\mathcal O(\Delta t, \Lambda (\Delta t)^{-2})$ and again can be neglected. Thus, the considerations mentioned in the paper do not seem to favor any particular choice of $a_n$ and $b_n$. However, to preserve time-reversal (anti)symmetry of $\widehat {\sigma^2}_{\mu \nu \alpha}$, it may be preferable to choose the $a_n$ symmetric and the $b_n$ antisymmetric.

Finally, I will discuss the appropriate generalization to multiple dimensions of the requirement that the measurement error be small compared to the typical displacement in a single time-step. The generalization is simple, but I feel it is worth spelling out explicitly. Symbolically, we have $\boldsymbol \Lambda \ll \widetilde {\mathbf C}_{\mathbf{vv}} := \langle \mathbf v \mathbf v^{\mathsf T} \rangle (\Delta t)^2$; this is to be interpreted by multiplying on the left by $\boldsymbol \xi^{\mathsf T}$ and on the right by $\boldsymbol \xi$ for any non-zero vector $\boldsymbol \xi$. Now consider $\boldsymbol \Lambda \widetilde {\mathbf C}_{\mathbf{vv}}^{-1}$; this is a second-rank mixed tensor whose Jordan normal form is independent of the choice of coordinates. (Assume that $\widetilde {\mathbf C}_{\mathbf{vv}}$ is non-singular; otherwise, the dynamics takes place in a subspace of lower dimension.) We may transform to a coordinate system in which $\widetilde {\mathbf C}_{\mathbf{vv}}$ is the identity. In this coordinate system, $\boldsymbol \Lambda \widetilde {\mathbf C}_{\mathbf{vv}}^{-1}$ is symmetric positive semidefinite; hence it is diagonalizable with non-negative eigenvalues. Let $\lambda$ be an eigenvalue of $\boldsymbol \Lambda \widetilde {\mathbf C}_{\mathbf{vv}}^{-1}$ with corresponding eigenvector $\boldsymbol \xi$, and let $\boldsymbol \xi' := \widetilde {\mathbf C}_{\mathbf{vv}}^{-1} \boldsymbol \xi$. Then $\boldsymbol \Lambda \boldsymbol \xi' = \lambda \widetilde {\mathbf C}_{\mathbf{vv}} \boldsymbol \xi'$, which upon multiplying on the left by ${\boldsymbol \xi'}^{\mathsf T}$ gives the requirement $\lambda \ll 1$. Note that we have implicitly assumed that the state variables $\mathbf x$ have a stationary probability distribution, so that $\langle \mathbf v \rangle = \mathbf 0$. If this is not the case, then $\mathbf v$ must be replaced by $\mathbf v - \langle \mathbf v \rangle$ in the criterion.

\section*{Acknowledgments}
The numerical calculations were performed with the help of Wolfram Mathematica. I would also like to thank D.\ B.\ Br\"{u}ckner for pointing out that the numerical simulations appear to be working properly, despite the errors in the written text.

\bibliographystyle{unsrt}
\bibliography{comment2}
\end{document}